\documentclass[pdftex,letterpaper,rmp,twocolumn,groupedaddress,floatfix]{revtex4}
\bibpunct{(}{)}{,}{n}{,}{,}
\usepackage[utf8]{inputenc}
\usepackage{graphicx}
\usepackage{amssymb}
\usepackage{fancybox}
\usepackage{amsmath}
\usepackage{color}
\usepackage{rotating}
\definecolor{darkgray}{rgb}{0.25,0.25,0.25}
\definecolor{darkred}{rgb}{0.89,0.10,0.11}
\definecolor{darkblue}{rgb}{0.12,0.39,0.62}
\usepackage{url}
\usepackage[pdftex,breaklinks=true,colorlinks=true,citecolor=black,linkcolor=black,menucolor=black,urlcolor=darkblue,pdfborder={1 0 0}]{hyperref}
\hypersetup{pdftitle={Multilevel compression of random walks on networks reveals hierarchical organization in large integrated systems},pdfauthor={Martin Rosvall and Carl T. Bergstrom, 2010}}

\newcommand{\T}{\rule{0pt}{3.6ex}}

\begin{document}
\makeatletter
\renewcommand\@biblabel[1]{#1.}
\makeatother
\title{Multilevel compression of random walks on networks reveals hierarchical organization in large integrated systems}
\author{M. Rosvall}
\email{martin.rosvall@physics.umu.se}
\homepage{http://www.tp.umu.se/~rosvall/}
\affiliation{Integrated Science Lab, Department of Physics, Umeå University, SE-901 87 Umeå, Sweden}
\author{C. T. Bergstrom}
\affiliation{Department of Biology, University of Washington, Seattle, WA 98195-1800}
\affiliation{Santa Fe Institute, 1399 Hyde Park Rd., Santa Fe, NM 87501}
\date{\today}

\begin{abstract}
To comprehend the hierarchical organization of large integrated systems, we introduce the hierarchical map equation, which reveals multilevel structures in networks. In this information-theoretic approach, we exploit the duality between compression and pattern detection; by compressing a description of a random walker as a proxy for real flow on a network, we find regularities in the network that induce this system-wide flow. Finding the shortest multilevel description of the random walker therefore gives us the best hierarchical clustering of the network --- the optimal number of levels and modular partition at each level --- with respect to the dynamics on the network. With a novel search algorithm, we extract and illustrate the rich multilevel organization of several large social and biological networks. For example, from the global air traffic network we uncover countries and continents, and from the pattern of scientific communication we reveal more than 100 scientific fields organized in four major disciplines: life sciences, physical sciences, ecology and earth sciences, and social sciences. In general, we find shallow hierarchical structures in globally interconnected systems, such as neural networks, and rich multilevel organizations in systems with highly separated regions, such as road networks.
\end{abstract}

\maketitle

\section*{Introduction}

Ever since Aristotle, organization and classification have been cornerstones of science. In network science \cite{albert2002,newmanSIAM}, categorization of nodes into modules with community-detection algorithms has proven indispensable to comprehending the structure of large integrated systems \cite{fortunato,girvan_newman,Ahn:2010p167}. But in real-world networks, the organization rarely is limited to two levels, and modular descriptions can only provide cross sections of much richer structures. For example, both biological and social systems are often characterized by hierarchical organization with submodules in modules over multiple scales \cite{winther2001,schlosser2004,ravasz2003hierarchical,clauset2008hsa,guimerahierarchy}.  

Several network clustering algorithms generate hierarchical trees, but few make more than a single cut 
through the dendrogram. To extract multiple levels of the network structure \cite{clauset2008hsa,guimerahierarchy,ronhovde2009multiresolution,kovacs2010community},
the common approach is to first generate a dendrogram or group nodes with one method and then determine the multiple cuts or the resolution thresholds with a different method. 
Moreover, these methods approach the problem of community detection by
inferring a model of an underlying generative process that created the
network. That is, they view the real network structure as a realization 
of a probabilistic process that creates links between groups of nodes
and try to identify the most likely underlying grouping. 
While this may be the appropriate strategy when one is
fundamentally interested in the modular nature of the dynamics by which a
given network was formed, it may not be optimal when one is more
interested in understanding the subsequent dynamics or behavior that
occur on the real network \cite{MullerLinow:2008p322}. 

In many real-world networks, directed and weighted links represent
the constraints that the structure of a network places on dynamical
processes taking place on this network. Networks thus often
represent literal or metaphorical flows: people surfing the web,
passengers traveling between airports, ideas spreading between
scientists, funds passing between banks, and so on. This flow through a system makes its components interdependent to varying extents.
The objective of our hierarchical clustering approach, therefore, is to reveal the multiple levels of interdependences between the nodes of a network with a single method.
That is, a method that does not require multiple external resolution parameters, but rather inherently reveals the natural multiple levels of the system. 

In this paper, we generalize the flow-based and information theoretic clustering method called the {\em map equation} \cite{RosvallBergstrom08,rosvall2009map} to uncover important multilevel structures and their relationships in networks. This generalization yields the {\em hierarchical map equation}, which provides a natural answer to three
questions: Into how many hierarchical levels is a given network
organized? How many modules are present at each level? And which nodes
are members of which modules?
Here we focus on hard partitions and flow of random walkers; we postpone the natural extension of this approach to overlapping partitions and generalized flows to a subsequent paper. We begin by briefly reviewing the map equation, and then introduce the hierarchical map equation, of which our earlier {\em two-level map equation} \cite{RosvallBergstrom08,rosvall2009map} can be seen as a special case. We then illustrate the mechanics of the hierarchical map equation, and extract and depict the hierarchical structure of several large-scale networks.
Finally, in the Materials and Methods section, we provide a detailed description and a performance test of our novel recursive search algorithm.

\section*{Results and Discussion}
\subsection*{The two-level map equation}
We have recently introduced the map equation to simplify and highlight important structures with respect to the dynamics on networks. This approach uses a random walk as a proxy for the real flow \cite{RosvallBergstrom08,rosvall2009map}, and exploits the duality between compressing a message and finding patterns in the structure that generates that message \cite{rissanen1978,grunwald}. To find the regularities that induce the dynamics on networks, the map equation measures, for a given network partition, the per-step average description length of a random walker moving along the (weighted and directed) links between the nodes of a network. By minimizing the map equation over all possible network partitions, we can reveal the structures that generate the flow on the network. 

\begin{figure*}[tb]
\centering
\includegraphics[width=0.9\textwidth]{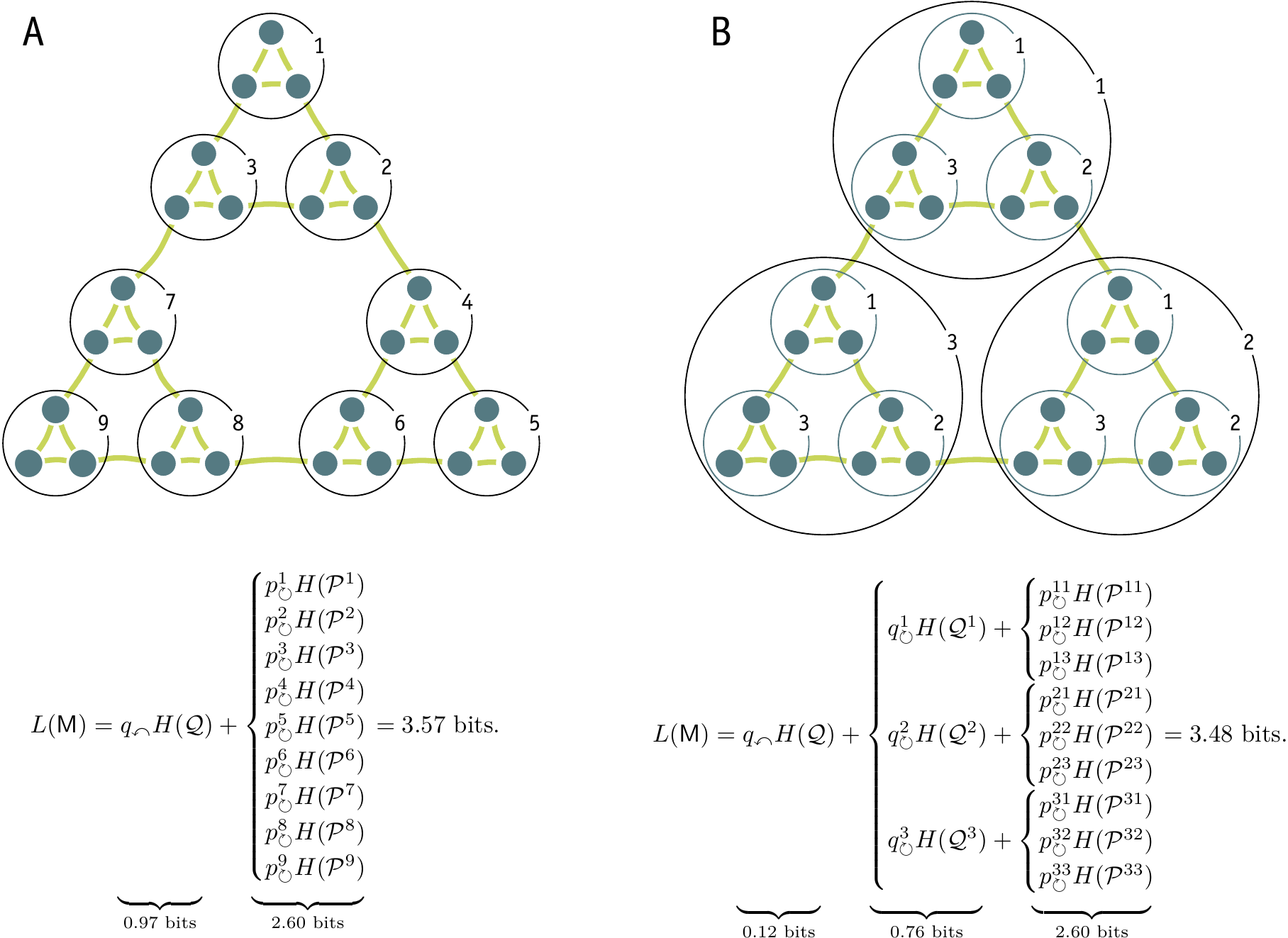}
\caption{
{\bf Minimizing the map equation over all network partitions gives an optimal clustering of the network with respect to the dynamics on the network.} Optimal two-level clustering is shown in A and hierarchical clustering is shown in B. The description length, which is 4.75 bits for an unpartitioned network, is the sum of the average length of codewords from the index codebook(s) and the module codebooks weighted by the rate of use of each codebook. For this undirected unweighted network with total degree 78, all rates can be calculated by counting links and normalizing: The codewords of the index codebook in A are used at relative rates
$\mathcal{Q} = \frac{3}{24},\frac{2}{24},\frac{3}{24},\frac{2}{24},\frac{3}{24},\frac{3}{24},\frac{3}{24},\frac{3}{24},\frac{2}{24}$ at a total rate $q_{\curvearrowleft} = \frac{24}{78}$ and, for example, the codewords of the first module codebook are used at relative rates $\mathcal{P}^1 = \frac{2}{10},\frac{3}{10},\frac{3}{10},\frac{2}{10}$ at a total rate $p_{\circlearrowright}^1 = \frac{10}{78}$ with contribution from the exit probability $q_{\curvearrowright}^1 = \frac{2}{78}$. The codewords of the smaller index codebooks in B are used at relative rates $\mathcal{Q} = \frac{2}{6},\frac{2}{6},\frac{2}{6}$ and $\mathcal{Q}^1 = \frac{2}{10},\frac{3}{10},\frac{3}{10},\frac{2}{10}$ at total rates $q_{\curvearrowleft} = \frac{6}{78}$ and $q_{\curvearrowleft}^{1} = \frac{10}{78}$. The fine-level modules of this hierarchical clustering coincide with the modules of the two-level clustering.
}
\label{fig1}
\end{figure*}

The map equation is designed to capitalize on the modular structure of
a network; the description length of the dynamics on the network can
be compressed if the network has localized regions in which small
groups of nodes have long persistence times. Compression is achieved by using multiple module codebooks with reused short codewords for different nodes in the network. To make the compressed description unambiguous, an index codebook distinguishes which module codebook is active. Specifically, for a module partition $\mathsf{M}$ of $n$ nodes $\alpha=1,2,\ldots,n$ into $m$ modules $i=1,2,\ldots,m$, the lower bound on the code length $L(\mathsf{M})$ is the sum of the average length of codewords for each codebook weighted by the rate of use of each codebook. Shannon's source coding theorem \cite{shannon48} states that, when we use $n$ codewords to describe the $n$ states of a random variable $X$ that occur with frequencies $p_i$, the average length of a codeword can be no less than the entropy of the random variable $X$ itself: $H(X) = -\sum_{1}^{n} p_i \log(p_i)$ (we measure code lengths in bits and take the logarithm in base 2). This gives us the map equation:
\begin{align}\label{map}
L(\mathsf{M}) = q_{\curvearrowright} H(\mathcal{Q}) + \sum_{i=1}^{m}p_{\circlearrowright}^iH(\mathcal{P}^i).
\end{align}
$H(\mathcal{Q})$ is the frequency-weighted average length of codewords in the index codebook, and $H(\mathcal{P}^i)$ is the frequency-weighted average length of codewords in module codebook $i$. Further, the entropy terms are weighted by the rate at which the codebooks are used. With $q_{i \curvearrowright}$ for the probability of exiting (and entering) module $i$, the index codebook is used at a rate $q_{\curvearrowright}=\sum_{i=1}^m q_{i \curvearrowright}$, which is the probability that the random walker switches modules on any given step. With $p_{\alpha}$ for the probability of visiting node $\alpha$, module codebook $i$ is used at a rate $p_{\circlearrowright}^i = \sum_{\alpha \in i} p_\alpha + q_{i \curvearrowright }$, the fraction of time the random walker spends in module $i$ plus the probability that she exits the module and the exit message is used. We have provided an interactive and dynamic visualization of the mechanics of the map equation here: \url{www.mapequation.org}.

Figure~\ref{fig1}A illustrates the partitioning obtained by using the two-level map equation. The 27-node example network is partitioned into nine modules, and the description length is theoretically 3.57 bits. For comparison, a single-module description of the network (one module codebook and no index codebook) has a lower bound of 4.75 bits.

When driven by a strong search algorithm, the map equation provides an efficient tool for revealing the modular
structure of networks \cite{LancichinettiFortunato09}. But many networks have important structures at multiple scales \cite{fortunato}, and the code structure of the two-level map equation cannot capitalize on these. For example, the network in Fig.~\ref{fig1}A is hierarchically organized with submodules within modules, but the
two-level map equation cannot simultaneously capitalize on both the
module and submodule levels of structure. It minimizes code
length by partitioning at the submodule level, revealing nine modules as shown
in Fig.~\ref{fig1}A. Additional potential for compression from the module
level structure goes untapped, and thus additional structure at the
module level goes unreported.

\subsection*{The hierarchical map equation}
To reveal pattern at multiple levels, we must generalize the coding
structure upon which the two-level map equation is based. Figure~\ref{fig1}B 
shows a hierarchical description of the network with not one but two
index codebooks, one for each level of hierarchy. With this code
structure, the description length can be reduced from the 3.57
bits required by the two-level map equation to 3.48 bits, because the
average description length to determine which of the nine module
codebooks is active has been reduced by 0.09 bits per step.
The extra codebook makes it possible
to exploit the fact that the fine-level modules are themselves organized into
larger modules: once a random walker enters one of the three  
larger modules, she tends to stay there for a long time.

\begin{figure*}[!ht]
\centering
\includegraphics[width=0.9\textwidth]{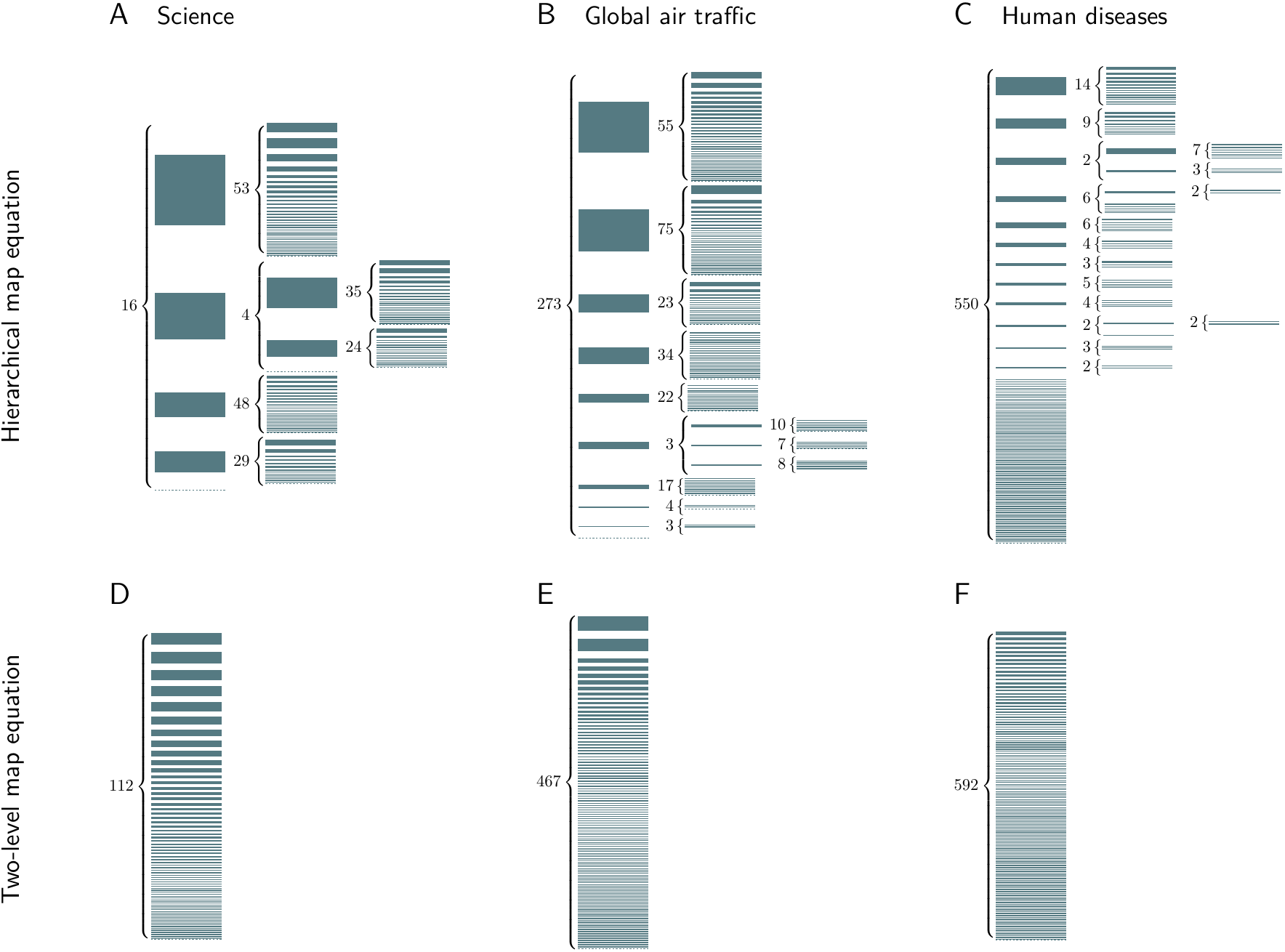}
\caption{
{\bf Multilevel organization in three real-world networks.} The bottom row illustrates structures that a two-level clustering can capture. The width of the horizontal lines represents the size of the modules and the number to the left of the braces gives the number of submodules within each module. For visual simplicity, we exclude submodules with less than 1 per mil of all flow. See Fig.~\ref{fig3} for a hierarchical map of science based on the journal citation network.
}
\label{fig2}
\end{figure*}

Broadly, in the hierarchical map equation we release the constraint of a single index codebook and allow for an arbitrary number of hierarchically nested index codebooks that specify movements between modules, submodules, subsubmodules, and so on, down to the finest modular level. 
Formally, for a hierarchical map $\mathsf{M}$ of $n$ nodes partitioned into $m$ modules, for which each module $i$ has a submap $\mathsf{M}^i$ with $m^i$ submodules, for which each submodule $ij$ has a submap $\mathsf{M}^{ij}$ with $m^{ij}$ submodules, and so on, the hierarchical map equation takes the form
\begin{align}
	L(\mathsf{M}) = q_{\curvearrowleft} H(\mathcal{Q}) + \sum_{i=1}^{m}L(\mathsf{M}^i),
\end{align}
with the description length of submap $\mathsf{M}^i$ at intermediate levels given by
\begin{align}
	L(\mathsf{M}^i) = q_{\circlearrowright}^i H(\mathcal{Q}^i) + \sum_{j=1}^{m^i}L(\mathsf{M}^{ij})
\end{align}
and at the finest modular level by
\begin{align}
	L(\mathsf{M}^{ij\ldots k}) = p_{\circlearrowright}^{ij\ldots k}H(\mathcal{P}^{ij\ldots k}).
\end{align}
At each submodule level, $q_{\circlearrowright}^{i}$ is the rate of codeword use for entering the $m_{i}$ submodules or exiting to a coarser level and $H(\mathcal{Q}^{i})$ is the frequency-weighted average length of the codewords in the subindex codebook. At the finest level, $p_{\circlearrowright}^{ij\ldots k}$ is the rate of codeword use for visiting nodes in submodules $ij\ldots k$ or exiting to a coarser level and $H(\mathcal{P}^{ij\ldots k})$ is the frequency weighted average length of the codewords in the submodule codebook. To find the hierarchical structure that best represents the structure with respect to flow, we seek the hierarchical partition of the network that minimizes the hierarchical map equation over all possible hierarchical partitions of the network (see Materials and Methods for a detailed description and a performance test of the algorithm). Figure~\ref{fig1}B illustrates the optimal hierarchical partition and the corresponding code structure for the example network.

\subsection*{Multilevel organization in real-world networks}
The hierarchical map equation can reveal rich multilevel organization in real-world networks. Figures \ref{fig2}A-C provide thumbnail illustrations of the hierarchical structure of the journal citation network of science \cite{rosvall2010mapping}, the global air traffic network \cite{guimera2005worldwide}, and the human disease network \cite{goh2007human}. For comparison, Figures \ref{fig2}D-F show the structure of each network as characterized by the two-level map equation.

\begin{figure*}[!ht]
\centering
\includegraphics[width=1.0\textwidth]{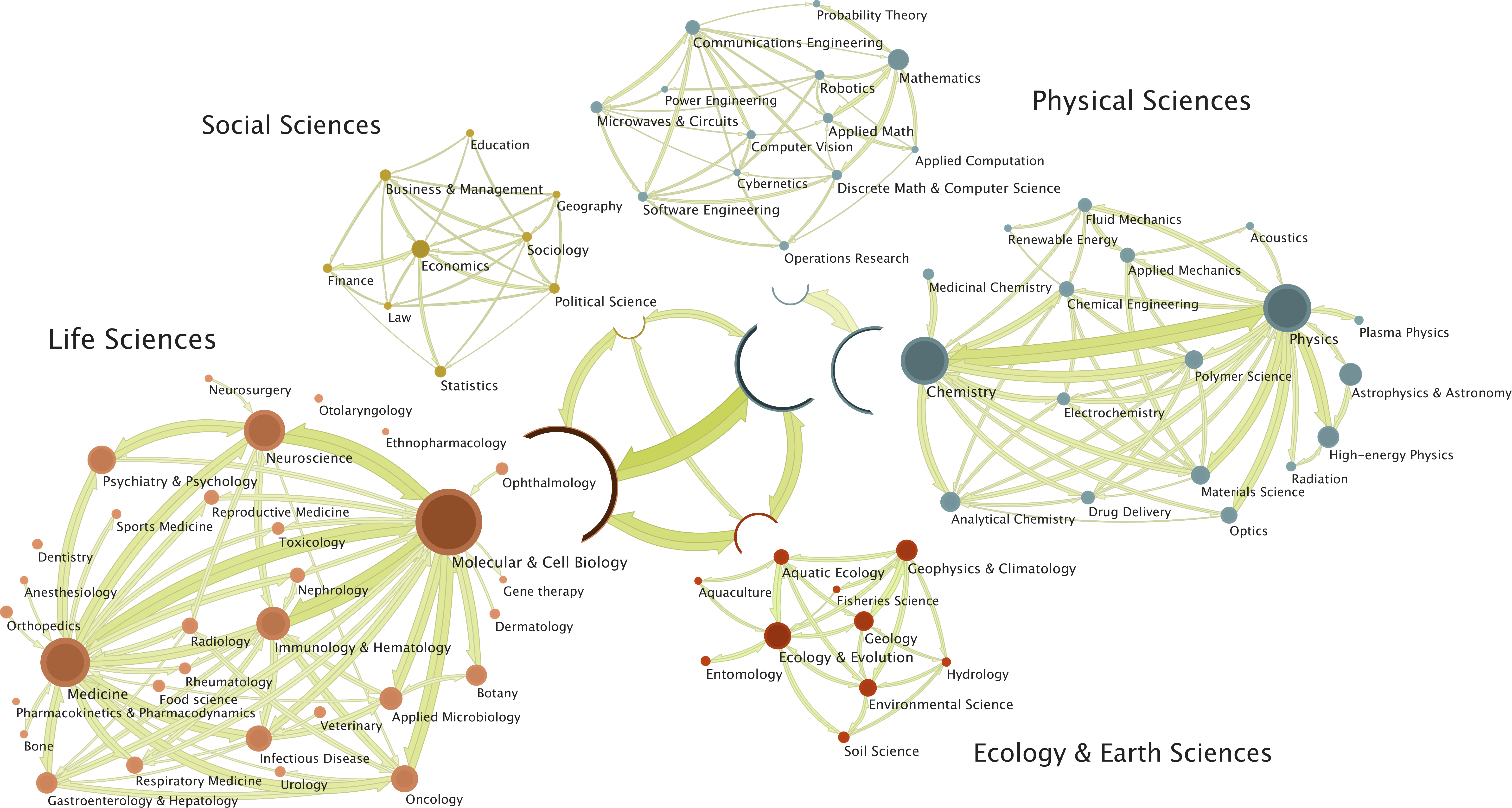}
\caption{
{\bf A hierarchical map of science.} We partitioned 7,940 journals connected by 9.2 million citations \cite{rosvall2010mapping} into four major disciplines, which we identified as life sciences, physical sciences, ecology and earth sciences, and social sciences. In physical sciences, we followed a second-level split into the areas of mathematics and of physics and chemistry. The size of the modules represents the fraction of time that a random surfer spends following citations in that field, and the arrows indicate flow volume between the fields. For visual simplicity, we exclude fields and arrows with low flow.
}
\label{fig3}
\end{figure*}

The journal citation network traces more than nine million citations among nearly 8,000 journals in the sciences and social sciences. From the pattern of citations, we reveal more than 100 scientific fields organized in four major disciplines: life sciences, physical sciences, ecology and earth sciences, and social sciences. The physical sciences are in turn organized into physics and chemistry, with 35 subfields, and mathematics, with 24 subfields (see Fig.~\ref{fig3}). 

In the global air traffic network, two cities are considered connected if a regularly scheduled commercial passenger flight travels between them. From the network of 3,883 cities connected by 14,142 links, the algorithm uncovers an overall organization of cities grouped in countries and countries grouped in continents. For example, the largest module comprises European and African cities arranged into 55 submodules; the second largest module comprises North and South American cities organized into 75 submodules. These submodules represent the Eastern US cities, the Western US cities, Mexican cities, and so on.

For the familiar networks of science and global air traffic, the organization revealed by the hierarchical map equation is intuitive and anticipated. But for the human disease network that connects diseases if they share common genes \cite{goh2007human}, the outcome is quite different. In the hierarchical partition of this network, the submodules contain class-related diseases, but only the largest module, which groups different cancers together, is compatible with any natural classification of diseases. We interpret this as an effect of missing data and a bias toward studies on oncogenes and other genes associated with cancer.

Beyond these three examples, many real-world networks have rich hierarchical structures. To
illustrate, we have used the generalized map equation to partition
twelve networks, ranging in size from hundreds to millions of nodes.
In Table \ref{table}, these networks are listed in descending order according
to the magnitude of the compression gained by using a multilevel
partitioning instead of a two-level partitioning.
In general, we find shallow hierarchical structures in globally interconnected systems and rich multilevel organizations in systems with highly separated regions.

\begin{table}[!ht]
\centering
\caption{{\bf The hierarchical organization of real-world networks.} For each multi-level classification of a network with $n$ nodes and $l$ links, we report the total number of modules $m$ together with the number of modules with more than one percent of all nodes, the per-node average depth $\langle d \rangle$, the per-node average size of the lowest-level module $\langle s_b \rangle$, and the compression gain over a two-level clustering $\Delta C$. The 16 networks are ordered by the compression gain, which provides information about how hierarchical the organization is.}
\centering
\begin{footnotesize}
\begin{tabular*}{\columnwidth}{@{\extracolsep{\fill}}lrrrrrrr}
\T Network & $n$ & $l$ & $m$ ({\tiny >$\frac{n}{100}$}) & $\langle d \rangle$ & $\langle s_b \rangle$ & $\Delta C$ \\
\hline
\T California roads \cite{leskovec2008community} & 2.0M & 5.5M & 0.45M (0) & 4.8 & 6.3 & 36\%\\ 
Google web$^{\textrm{d}}$ \cite{leskovec2008community} & 0.74M & 5.1M & 73k (34) & 4.5 & 0.67k & 16\%\\  
Stanford web$^{\textrm{d}}$ \cite{leskovec2008community} & 0.28M & 2.3M & 35k (41) & 4.9 & 0.20k & 15\%\\
Call graph$^{\textrm{wd}}$ \cite{ourdata} & 2.5k & 7.2k & 0.91k (53) & 4.7 & 8.3 & 8.0\%\\ 
Coauthorships$^{\textrm{w}}$ \cite{ourdata} & 0.55k & 1.3k & 94 (55) & 3.3 & 9.5 & 5.7\% \\ 
Human diseases$^{\textrm{w}}$ \cite{goh2007human} & 1.3k & 1.5k & 0.62k (23) & 2.5 & 5.4 & 4.5\% \\
Global air traffic \cite{guimera2005worldwide} & 3.9k & 14k & 0.53k (27) & 3.0 & 46 & 2.3\% \\
Stockholm roads$^{\textrm{*}}$ \cite{rosvall2005networks} & 11k & 23k & 1.0k (19) & 3.1 & 16 & 1.7\% \\ 
Journal citations$^{\textrm{wd}}$ \cite{rosvall2010mapping} & 7.9k & 1.1M & 0.21k (32) & 3.3 & 0.16k & 1.6\% \\ 
Political blogs$^{\textrm{wd}}$ \cite{ourdata} & 1.1k & 13k & 0.28k (13) & 2.8 & 75 & 0.0\% \\ 
US airports$^{\textrm{wd}}$ \cite{USairports} & 0.50k & 18k & 14 (9) & 2.0 & 0.14k & 0.0\% \\
\textit{C. Elegans} brain$^{\textrm{wd}}$ \cite{watts1998collective} & 0.30k & 2.3k & 22 (18) & 2.0 & 35 & 0.0\% \\ 
\hline
\multicolumn{8}{l}{\scriptsize ${^{\textrm{w}}}$Weighted links. ${^{\textrm{d}}}$Directed links.}\\ 
\multicolumn{8}{l}{\scriptsize${^{\textrm{*}}}$Dual representation with roads as nodes and intersections as edges.}\\
\end{tabular*}
\end{footnotesize}
\label{table}
\end{table}

The network with the
highest compression gain --- i.e., the network with the greatest degree
of nested hierarchical structure --- is the California road network \cite{leskovec2008community}. The geographical constraints of the road network prevent shortcuts between different and remote parts of the network. As a result, the organization is distinct down to the very many small bottom modules. The web graphs have the next greatest compression gain. They are as deep as the road network, but without physical constraints, different parts of the web are presumably more interconnected. The lowest-level are on average larger, and the flow between different large-scale regions reduces the compression gain.

In the other extreme in Table \ref{table} are the \textit{C. Elegans} brain network \cite{watts1998collective} and the weighted and directed network of US air travel passengers \cite{USairports}, which were best compressed by two-level descriptions. The many links between different regions at a global scale of these networks maintain high connectivity and short distances, and prevent further gain from a multilevel description. For the same reason, the dual road network of Stockholm \cite{rosvall2005networks}, with roads as nodes and intersections as edges, has a less pronounced multilevel structure than the road network of California \cite{leskovec2008community}, with intersections as nodes and roads as links, and the different representations overshadow differences in the actual road layouts. For example, a main road that intersects with many streets in several suburbs forms a hub that connects suburban streets in the dual representation. Therefore, the gain from a deep multilevel description is lost in the dual representation, which suppresses distances and makes the network more interconnected. When comparing the hierarchical depth between the road network of California and the dual road network of Stockholm, the range of the networks also plays an important role. Both networks represent streets in neighborhoods in suburbs, but the road network of California also includes the additional level of multiple cities. In this way, and because the number of nodes in a network quickly grows for every additional level of nested modules, there is a general trend that the hierarchical depth increases with network size in Table \ref{table}.

Figure~\ref{fig2} and Table \ref{table} summarize the extent of hierarchical structure found in several large networks, but they provide no information about the relationships among the modules at any given level. To comprehend the dynamics of a system, we must capture both its hierarchical structure and the connections among modules at all levels of structure. Because the hierarchical map equation naturally balances the persistence times in modules and the flow between modules when it exploits the regularities in patterns of movement on a network, both are intrinsic to our approach. 
In Fig.~\ref{fig3}, we illustrate the relationships among modules in a hierarchical map of science. The multilevel map highlights and simplifies the citation flow between the major disciplines. At the same time, it summarizes the flows between fields that integrate those fields into larger disciplinary areas; for example, the arrows indicate the flows among the fields composing the social sciences. If a researcher would make a random walk in the scholarly literature by reading a paper and following a random citation to a new paper, she would spend 54 percent of her time reading journals in the life sciences, 33 percent in the physical sciences, 8 percent in the ecology and earth sciences, and 4 percent in the social sciences. The disciplines are well defined with long persistence times; only around one percent of the time would she follow a citation across discipline boundaries, the traversal from the physical sciences to the life sciences being the most common of these.

Using the fundamental mathematics of information theory to exploit the duality between compression and pattern detection, we have shown how to reveal the multilevel organization of networks. Combined with powerful visualizations, the hierarchical map equation provides a useful tool to comprehend the hierarchical organization of large multiscale social and biological systems. Here we have focused on hard partitions and the flow of random walkers, but in a subsequent paper we will demonstrate the natural extension of the map equation to overlapping partitions and generalized flows. In short, we can capitalize on overlapping structures by modifying the code structure and releasing the constraint that a node can only belong to one module codebook. Because the codelength only depends on the rates of node visits and module transitions, the map equation framework is agnostic to the origin of the flow. Therefore, we can comprehend the organization in real systems for which a random walker is not a good proxy for flow through the system, by using a different model of flow or by directly measuring the real flow.

\section*{Materials and Methods}
Here we provide a detailed description of the mathematics of the hierarchical map equation and outline the stochastic and recursive algorithm we have developed to search for the hierarchical partition of a network that minimizes the hierarchical map equation. We also describe how we quantify the performance of our method with the relative mutual information of module and submodule assignments between the benchmark networks and the hierarchical clustering generated by the algorithm. 

\subsection*{The hierarchical map equation}
The hierarchical partitioning algorithm builds on the fast stochastic search algorithm presented in ref.\ \cite{rosvall2009map}, with two major differences. 
First, to explore multilevel solutions, the algorithm recursively tries to add extra index codebooks both at coarser and finer levels. Sometimes movements between modules can be further compressed by adding one or more coarser index codebooks and sometimes movements within modules can be further compressed by adding one or more finer index codebooks. In its search for the optimal hierarchical partitioning, the algorithm successively increases and decreases the depth of different branches of the multilevel code structure.
Second, to reduce the small cohesive effect of random teleportation, the map equation only measures the description length of steps following links and not the steps associated with random teleportation. In this way, the resolution increases slightly and the algorithm can better detect less-separated modules or submodules. The code is available here: \url{http://www.tp.umu.se/~rosvall/code.html}. Below we explain how we have implemented these differences.

To exclude random teleportation steps from the description length of directed networks, we first calculate the ergodic node visit frequencies $p_{\alpha}$ for $\alpha=1,\ldots,n$ with random teleportation at rate $\tau = 0.15$ as before. Then, for every node $\alpha$ and for all its outgoing links with relative weight $w_{\alpha\beta}$ to node $\beta$, we calculate the probability that the random surfer does not teleport but rather follows a link in a given step:
\begin{align}
	q_{\alpha \curvearrowright \beta} = (1-\tau) p_{\alpha} w_{\alpha\beta}.
\end{align}
Note that the in- and outflow no longer need to be equal, as in the ergodic case. 
Finally, we update the node visit frequencies to exclude the contribution from random teleportation:
\begin{align}
	p_{\alpha} = \sum_{\beta}^{\alpha} q_{\beta \curvearrowright \alpha}.
\end{align}

For a given hierarchical network partition, the hierarchical map equation measures the per-step average minimal information necessary to track a random walker's movements along links on a network. Sometimes the random walker stays within the same finest-level submodule, and sometimes she moves up and down one or more levels in the hierarchy. At the coarsest level, the description length measures the information necessary to determine which coarsest-level module the random walker enters, weighted by how often such movements happen. The relative rate of codeword use is $\mathcal{Q} = \{q_{\curvearrowleft}^i/q_{\curvearrowleft}\} =
q_{\curvearrowleft}^1/q_{\curvearrowleft}, q_{\curvearrowleft}^2/q_{\curvearrowleft}, \ldots, q_{\curvearrowleft}^m/q_{\curvearrowleft}$, where 
\begin{align}
q_{\curvearrowleft} = \sum_{i=1}^m q_{\curvearrowleft}^i
\end{align} 
is the per-step average flow into the modules and the total codeword use at the coarsest level. The Shannon information of movements at the coarsest level --- weighted by the total use --- is therefore
\begin{align}
q_{\curvearrowleft}H(\mathcal{Q}) = q_{\curvearrowleft}\left( -\sum_{i=1}^m \frac{q_{\curvearrowleft}^i}{q_{\curvearrowleft}}\log \frac{q_{\curvearrowleft}^i}{q_{\curvearrowleft}}\right).
\end{align}
At intermediate levels, to measure the contribution to the total codelength in submodule $i$, it is sufficient to aggregate the flow associated with movements to coarser levels $q_{\curvearrowright}^i$ and flow that is associated with movements into the $m^{i}$ finer levels of the hierarchy $\{q_{\curvearrowleft}^{ij}\}$. The relative rate of codeword use is $\mathcal{Q}^i=q_{\curvearrowright}^i/q_{\circlearrowright}^i, q_{\curvearrowleft}^{i1}/q_{\circlearrowright}^i, \ldots, q_{\curvearrowleft}^{im^{i}}/q_{\circlearrowright}^i$, where 
\begin{align}
q_{\circlearrowright}^i = q_{\curvearrowright}^i + \sum_{j=1}^{m^{i}}q_{\curvearrowleft}^{ij}	
\end{align}
is the total codeword use. The Shannon information of movements in this submodule, weighted by how often the code is used, is therefore
\begin{align}
	q_{\circlearrowright}^iH(\mathcal{Q}^i) = q_{\circlearrowright}^i\left(-\frac{q_{\curvearrowright}^i}{q_{\circlearrowright}^i}\log\frac{q_{\curvearrowright}^i}{q_{\circlearrowright}^i} -\sum_{j=1}^{m^{i}}\frac{q_{\curvearrowleft}^{ij}}{q_{\circlearrowright}^i}\log\frac{q_{\curvearrowleft}^{ij}}{q_{\circlearrowright}^i}  \right).
\end{align}
At the finest levels, nodes rather than submodules are visited and the relative rate of codeword use is $\mathcal{P}^{ij\ldots k} = q_{\curvearrowright}^{ij\ldots k}/p_{\circlearrowright}^{ij\ldots k}, \{p_{\alpha \in {ij\ldots k}}/p_{\circlearrowright}^{ij\ldots k}\}$, where 
\begin{align}
p_{\circlearrowright}^{ij\ldots k} = q_{\curvearrowright}^{ij\ldots k} + \sum_{\alpha \in {ij\ldots k}}p_{\alpha}
\end{align}
is the total codeword use. The Shannon information of movements at the finest level weighted by the total use of the code therefore is
\begin{align}
&p_{\circlearrowright}^{ij\ldots k}H(\mathcal{P}^{ij\ldots k}) = \nonumber \\
&p_{\circlearrowright}^{ij\ldots k}\left(-\frac{q_{\curvearrowright}^{ij\ldots k}}{p_{\circlearrowright}^{ij\ldots k}}\log \frac{q_{\curvearrowright}^{ij\ldots k}}{p_{\circlearrowright}^{ij\ldots k}} - \sum_{\alpha \in {ij\ldots k}} \frac{p_{\alpha}}{p_{\circlearrowright}^{ij\ldots k}}\log \frac{p_{\alpha}}{p_{\circlearrowright}^{ij\ldots k}} \right).
\end{align}

Adding the contribution from every module at all levels gives the total description length, which is quantified by the hierarchical map equation. For a hierarchical map $\mathsf{M}$ of $n$ nodes partitioned into $m$ modules, for which each module $i$ has a submap $\mathsf{M}^i$ with $m^i$ submodules, for which each submodule $ij$ has a submap $\mathsf{M}^{ij}$ with $m^{ij}$ submodules, and so on, the hierarchical map equation takes the form
\begin{align}
	L(\mathsf{M}) = q_{\curvearrowleft} H(\mathcal{Q}) + \sum_{i=1}^{m}L(\mathsf{M}^i),
\end{align}
with the description length of submap $\mathsf{M}^i$ at intermediate levels given by
\begin{align}
	L(\mathsf{M}^i) = q_{\circlearrowright}^i H(\mathcal{Q}^i) + \sum_{j=1}^{m^i}L(\mathsf{M}^{ij})
\end{align}
and at the finest modular level by
\begin{align}
	L(\mathsf{M}^{ij\ldots k}) = p_{\circlearrowright}^{ij\ldots k}H(\mathcal{P}^{ij\ldots k}).
\end{align}

\subsection*{Fast stochastic and recursive search algorithm}
The hierarchical map equation measures the per-step average code length necessary to describe a random walker's link movements on a network, given a hierarchical network partition, but the challenge is to find the partition that minimizes the description length. Into how many hierarchical levels should a given network be partitioned? How many modules should each level have? And which nodes should be members of which modules? 

We have generalized our search algorithm for the two-level map equation to recursively search for multilevel solutions. The recursive search operates on a module at any level; this can be all the nodes in the entire network, or a few nodes at the finest level. For a given module, the algorithm first generates submodules if this gives a shorter description length. If not, the recursive search does not go further down this branch. 
But if adding submodules gives a shorter description length, the algorithm tests if movements within the module can be further compressed by additional index codebooks. Further compression can be achieved both by adding one or more coarser codebooks to compress movements between submodules or by adding one or more finer index codebooks to compress movements within submodules. To test for all combinations, the algorithm calls itself recursively, both operating on the network formed by the submodules and on the networks formed by the nodes within every submodule. In this way, the algorithm successively increases and decreases the depth of different branches of the multilevel code structure in its search for the optimal hierarchical partitioning.
For every split of a module into submodules, we use the search algorithm detailed in ref.\ \cite{rosvall2009map} and described again here.

Any greedy (fast but inaccurate) or Monte Carlo-based (accurate but slow) approach can be used to minimize the map equation. To provide a good balance between the two extremes, we developed a fast stochastic and recursive search algorithm, implemented it in C++, and made it available online both for directed and undirected weighted networks \cite{mapcode}. As a reference, the new algorithm is as fast as the previous high-speed algorithms (the greedy search presented in the supporting appendix of ref.~\cite{RosvallBergstrom08}), which were based on the method introduced in ref.~\cite{clauset-2004-70} and refined in ref.~\cite{wakita}. At the same time, it is also more accurate than our previous high-accuracy algorithm (a simulated annealing approach) presented in the same supporting appendix.

The core of the algorithm follows closely the method presented in ref.~\cite{blondel2008}: neighboring nodes are joined into modules, which subsequently are joined into supermodules, and so on. First, each node is assigned to its own module. Then, in random sequential order, each node is moved to the neighboring module that results in the largest decrease of the map equation. If no move results in a decrease of the map equation, the node stays in its original module. This procedure is repeated, each time in a new random sequential order, until no move generates a decrease of the map equation. Now the network is rebuilt, with the modules of the last level forming the nodes at this level, and, exactly as at the previous level, the nodes are joined into modules. This hierarchical rebuilding of the network is repeated until the map equation cannot be reduced further. Except for the random sequence order, this is the algorithm described in ref.~\cite{blondel2008}.

With this algorithm, a fairly good clustering of the network can be found in a very short time. Let us call this the core algorithm and see how it can be improved. The nodes assigned to the same module are forced to move jointly when the network is rebuilt. As a result, what was an optimal move early in the algorithm might have the opposite effect later in the algorithm. Because two or more modules that merge together and form one single module when the network is rebuilt can never be separated again in this algorithm, the accuracy can be improved by breaking the modules of the final state of the core algorithm in either of the two following ways:

\begin{itemize}
\item[] \emph{Submodule movements.} First, each cluster is treated as a network on its own and the main algorithm is applied to this network. This procedure generates one or more submodules for each module. Then all submodules are moved back to their respective modules of the previous step. At this stage, with the same partition as in the previous step but with each submodule being freely movable between the modules, the main algorithm is re-applied.

\item[] \emph{Single-node movements.} First, each node is re-assigned to be the sole member of its own module, in order to allow for single-node movements. Then all nodes are moved back to their respective modules of the previous step. At this stage, with the same partition as in the previous step but with each single node being freely movable between the modules, the main algorithm is re-applied.
\end{itemize}

In practice, we repeat the two extensions to the core algorithm in sequence and as long as the clustering is improved. Moreover, we apply the submodule movements recursively. That is, to find the submodules to be moved, the algorithm first splits the submodules into subsubmodules, subsubsubmodules, and so on until no further splits are possible. Finally, because the algorithm is stochastic and fast, we can restart the algorithm from scratch every time the clustering cannot be improved further and the algorithm stops. The implementation is straightforward and, by repeating the search more than once, 100 times or more if possible, the final partition is less likely to correspond to a local minimum. For each iteration, we record the clustering if the description length is shorter than the previous shortest description length. In practice, for networks with on the order of 10,000 nodes and 1,000,000 directed and weighted links, each iteration takes a few seconds on a modern laptop.

\subsection*{Performance test of the hierarchical map equation}
To test the performance of our algorithm, we used the benchmark paradigm developed by Lancichinetti and Fortunato \cite{LancichinettiFortunato09}. They have provided an extension of their algorithm to generate benchmark networks with an extra submodular level and made it available here: \url{http://sites.google.com/site/santofortunato/inthepress2}. But before detailing the performance test, we follow the reasoning in ref.\ \cite{LancichinettiFortunato09} and provide an approximate relationship between a well-defined hierarchical structure and the coarse- and fine-level mixing parameters.

\begin{figure}[!ht]
\centering
\includegraphics[width=0.9\columnwidth]{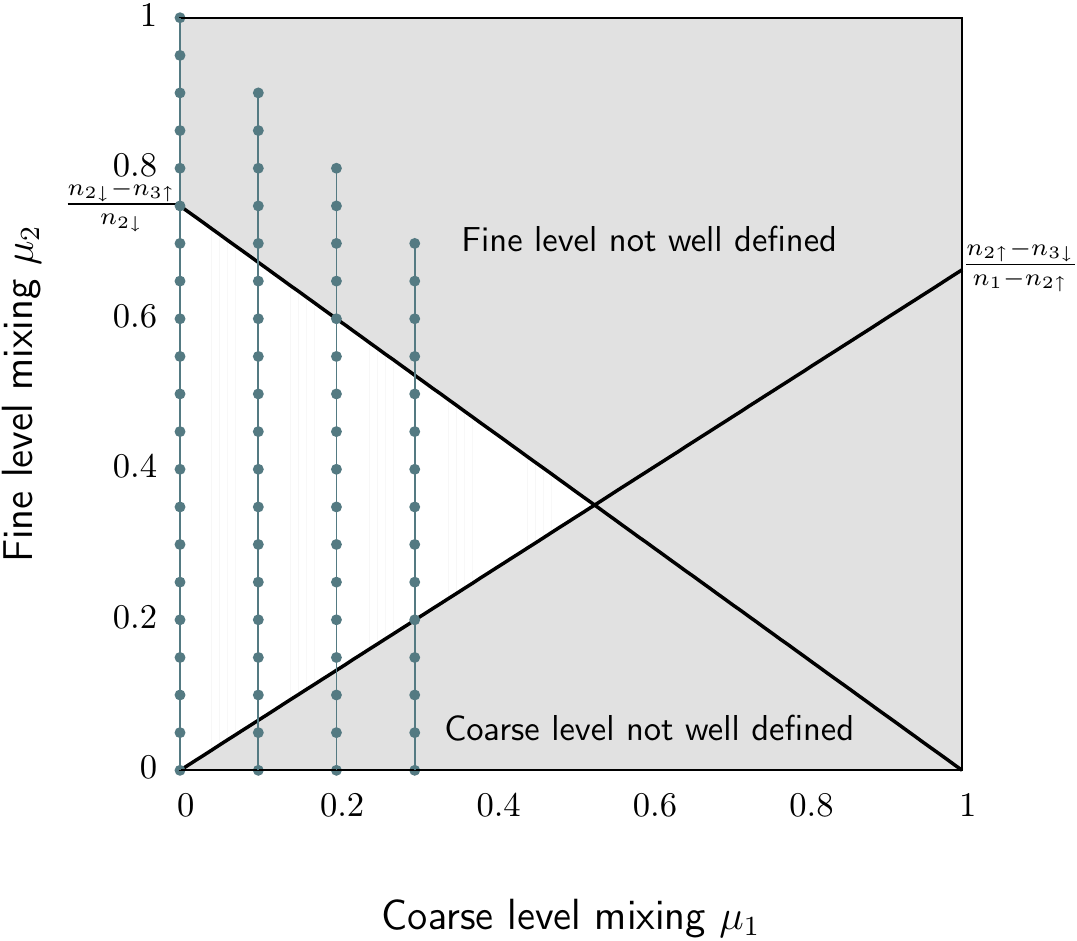}
\caption{
{\bf The range of mixing parameters that give a well-defined three-level hierarchical structure for the benchmark networks in the paper.} The networks have $n_1=10,000$ nodes, coarse-level module sizes between $n_{2\downarrow}=400$ and $n_{2\uparrow}=4,000$ nodes, and fine-level module sizes between $n_{3\downarrow}=10$ and $n_{3\uparrow}=100$ nodes. The connected points illustrate the sets of mixing parameters we present in the paper.
}
\label{fig4}
\end{figure}

From a topological point of view, a three-level hierarchical structure is well defined if
\begin{align}\label{eq1}
	p_3 > p_2 > p_1,
\end{align}
where $p_3$ is the probability that a random link connects two nodes in the same fine-level module, $p_2$ is the probability that it connects two nodes in different fine-level modules but the same coarse-level module, and $p_1$ is the probability that it connects two nodes in different coarse-level modules.  We can estimate these probabilities, given the expected number of links a node $i$ shares with nodes within the same fine-level module $k_i^3$, with nodes within the same coarse-level module but different fine-level modules $k_i^2$, and with nodes in other coarse-level modules $k_i^1$, We do this by approximating the number of available links within the same module to $n_3\langle k \rangle$, where $n_3$ is the number of nodes in the fine-level module and $\langle k \rangle$ is the average degree of nodes in the network. The corresponding approximation for within-coarse-level modules is $(n_2-n_3)\langle k \rangle$, where $n_2$ is the number of nodes in the coarse-level module. The approximation for available links in other coarse-level modules is $(n_1-n_2)\langle k \rangle$, where $n_1$ is the number of nodes in the full network. Now we have
\begin{align}\label{eq2}
	p_3 &\sim \frac{k_i^3}{n_3\langle k \rangle}\\
	p_2 &\sim \frac{k_i^2}{(n_2-n_3)\langle k \rangle}\\
	p_1 &\sim \frac{k_i^1}{(n_1-n_2)\langle k \rangle}.
\end{align}
The mixing parameters $\mu_1$ and $\mu_2$ are defined as follows:
\begin{align}
	1-\mu_2-\mu_1 & = \frac{k_i^3}{k_i^3+k_i^2+k_i^1}\\
	\mu_2 & = \frac{k_i^2}{k_i^3+k_i^2+k_i^1}\\
	\mu_1 & = \frac{k_i^1}{k_i^3+k_i^2+k_i^1}\label{eq3},
\end{align}
such that nodes share on average a fraction $\mu_1$ of their links with nodes in other modules, a fraction $\mu_2$ of their links with nodes in other submodules, and the remaining fraction $1-\mu_1-\mu_2$ of their links with nodes in the same submodule. Now we have the information to determine where the full hierarchical structure is well defined. Combining eqs.\ (\ref{eq1}-\ref{eq3}) yields the relationship
\begin{align}
	\frac{1-\mu_2-\mu_1}{n_3} > \frac{\mu_2}{n_2-n_3} > \frac{\mu_1}{n_1-n_2}.
\end{align}
The two inequalities correspond to two lines in the $\mu_1$-$\mu_2$ plane, determined by the extreme values of $n_3$, $n_2$, and $n_1$. For a well-defined three-level hierarchical structure, $\mu_2$ must be larger than
\begin{align}
	\frac{n_{2\uparrow}-n_{3\downarrow}}{n_1-n_{2\uparrow}}\mu_1
\end{align}
and smaller than
\begin{align}
	\frac{n_{2\downarrow}-n_{3\uparrow}}{n_{2\downarrow}}(1-\mu_1).
\end{align}
Here $n_{3\downarrow}$ is the smallest number and $n_{3\uparrow}$ the largest number of nodes a fine-level module can have, with the same notation for the coarse-level modules. Figure~\ref{fig4} shows the range of mixing parameters that correspond to a well-defined three-level hierarchical structure, for the values we have used in the benchmark test.

To quantify the performance of our method, we use the relative mutual information \cite{danon2005} and measure how much we learn about the true benchmark partitions by studying the inferred partitions that we get by applying the hierarchical map equation. We independently compare the coarse and fine levels of the benchmark networks with the multilevel partitioning inferred by the map equation. That is, we compare the first-level modules of the benchmark networks with the first-level modules of the inferred modules and the second-level submodules of the benchmark networks with the finest-level submodules of the inferred modules. Note that with this approach, the finest-level submodules do not need to be at the second level in the inferred structure. Therefore, we also measured the per-node average depth of the hierarchy to pick up information about how many levels were detected.

To calculate the relative mutual information, we label every node by its module number. In this way, picking a random node and reading off its module number corresponds to sampling from the discrete random variable $X$ with probability distribution $P(X)=n_1/n,n_2/n,\ldots,n_m/n$, where $n$ is the number of nodes, $n_x$ is the number of nodes in module $x$, and $m$ is the number of modules. The average information necessary to describe the random variable, the Shannon information of $X$, is accordingly
\begin{align}
H(X) = -\sum_{x} \frac{n_x}{n}\log \frac{n_x}{n}.
\end{align}
With $X$ for the benchmark partition, $Y$ for the algorithm partition, and $n_{xy}$ for the number of nodes that are jointly partitioned in module $x$ and module $y$, the mutual information is
\begin{align}
I(X;Y) = -\sum_{x,y} \frac{n_{xy}}{n}\log \frac{n\, n_{xy}}{n_x\, n_y}.
\end{align}

Finally, the normalized mutual information \cite{danon2005} with a range between $0$ for independent partitions and $1$ for identical partitions is 
\begin{align}
R(X;Y) = \frac{2I(X;Y)}{H(X)+H(Y)}.
\end{align}

\begin{figure}[!ht]
\centering
\includegraphics[width=1.0\columnwidth]{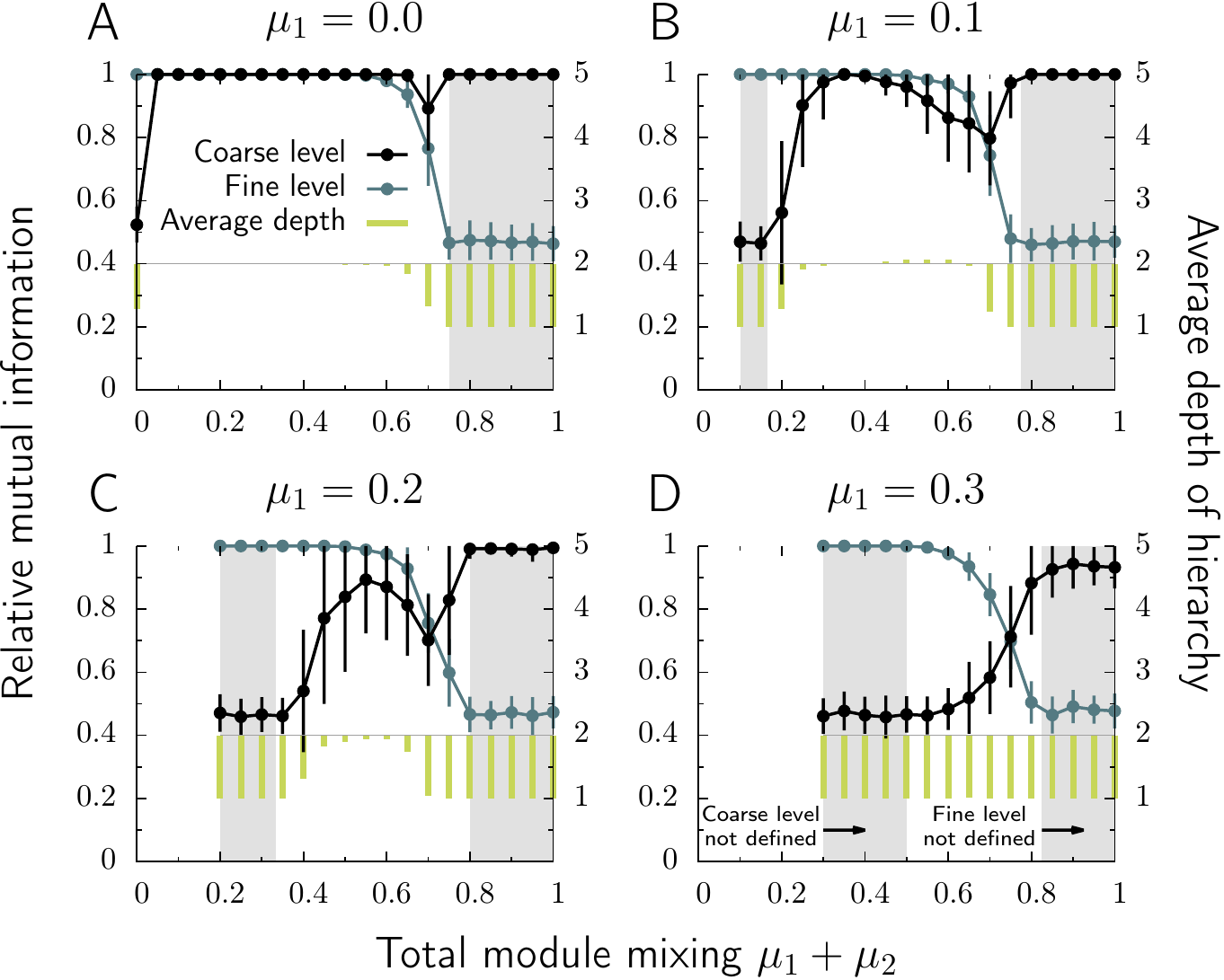}
\caption{
{\bf Hierarchical benchmark test.} Figures A-D show how well the algorithm reveals the three-level organization of the hierarchical benchmark networks with 10,000 nodes and 100,000 links. The nodes share a fraction $\mu_1$ of their links with nodes in other coarse-level modules and a fraction $\mu_2$ of their links with nodes in other fine-level modules. Every data point represents the average value of 100 measures. 
}
\label{fig5}
\end{figure}

We used scale-free networks (exponent -2) with 10,000 nodes, average degree 20, and maximum degree 100, and let the module sizes vary between 400 and 4,000 nodes and the submodule sizes between 10 and 100 nodes, both with a scale-free size distribution (exponent -1). Figure~\ref{fig5} shows the result of the benchmark test. The performance is excellent as long as the hierarchical organization is well defined and nodes have strictly more links within than between fine-level modules and more links within than between coarse-level modules; otherwise, the well-defined range is too narrow. Because of fluctuations in the benchmark networks, the levels interweave close to the limits of well-defined modules and the algorithm can only extract the fine-level modules. Overall, the results are on par with what we have obtained for two-level benchmark networks \cite{LancichinettiFortunato09}.

\begin{acknowledgments}
We are grateful to Andrea Lancichinetti for providing the code used to perform the multilevel benchmark test, to Jevin West for processing the data used to construct the hierarchical map in Fig.~\ref{fig3}, and to Alcides Viamontes Esquivel for compiling the call graph and the political blogs in Table \ref{table}. MR was supported by the Swedish Research Council grant 2009-5344. CTB was supported by NSF grant SBE-0915005 and by US NIGMS MIDAS Center for Communicable Disease Dynamics 1U54GM088588 at Harvard University.
\end{acknowledgments}


\begin{thebibliography}{10}
\bibitem{ourdata}
(year?)
\newblock We have compiled the call graph from the cross-platform library GLib,
  the coauthorship network from the references in refs.\ 1-3, and the political
  blogs from data provided by Twingly.com.
\end{thebibliography}

\end{document}